# Highly vibration-resistant sub-Hertz ultra-stable laser passing over 1700 km transport test

Dongdong Jiao [a,b,c], Xue Deng [a,b,c], Jing Gao [a,b], Linbo Zhang [a,b], Guanjun Xu [a,b,c*], Tao Liu [a,b,*], Ruifang Dong [a,b,*], Shougang Zhang [a,b]

[a] *National Time Service Center, Chinese Academy of Sciences, Xi'an, 710600, China*

[b] *Key Laboratory of Time and Frequency Standards, Chinese Academy of Sciences, Xi'an, 710600, China*

[c] *Both the authors contributed equally to this work*

*Correspondence authors：xuguanjun@ntsc.ac.cn; taoliu@ntsc.ac.cn; dongruifang@ntsc.ac.cn

**Abstract:** We present a compact ultra-stable laser at 1550 nm with a line-width of 0.57 Hz, achieved using a 50 mm length cubic optical cavity. With the help of elaborate investigation of the vibration behavior of the cavity in three directions, its physical structure was optimized for significantly improved insensitivity to vibration by increasing the first-order resonance frequency almost twice that of the previous setup. The anti-vibration capability of the ultra-stable laser was then verified in the long-distance highway transport test over more than 1700 km, which shows nearly unchanged performance compared to that before transport. A fractional frequency instability of ~$2.6 \times 10^{-15}$ was observed at 1–10 s, approaching the thermal noise floor. To our knowledge, this is the longest-distance transport vibration test for a practical transportable laser to date. The presented system will act as the local optical oscillator for high-precision remote optical frequency transfer, and the design strategy is valuable for space-borne laser applications.

1. Introduction

Optical atomic clocks have reached unprecedented fractional frequency instabilities below the $10^{-18}$ level [1-5], which have made them an indispensable experimental tool in numerous applications, such as fundamental physics tests [6-10], relativistic geodesy [11-14], and redefinition of the unit "second" [15,16]. However, the optical atomic clocks distributed in different regions will be nontrivially influenced by the gravitational redshift [17, 18], resulting in non-unified standards for these high precision applications. Optical frequency transfer over fiber links is found an effective way to address this problem, as excellent performance can be achieved for the comparison and synchronization of optical atomic clocks [8-12, 19-21]. Up to now, many countries have launched the projects of establishing fiber-optic time-frequency service networks, such as the clock network services (CLONETS) [22, 23] in Europe and the highly accurate ground-based time service system (HAGTSS) in China [24,25]. Within the projects, the optical frequency transfer is essential. In practical systems, to prevent the transferred optical frequency stability degradation, it is necessary to utilize narrow-linewidth ultra-stable lasers as the local oscillators [20, 21, 26]. For field applications, especially in harsh situations with significant vibrations, the robustness of the ultra-stable lasers is also crucial.

The ultra-stable laser is typically achieved by frequency stabilizing to the resonance of a high finesse Fabry-Pérot (F-P) optical cavity using the Pound-Drever-Hall (PDH) technique, excellent performance of sub-Hertz line-width and $10^{-16}$ frequency instability has been achieved in the laboratory environments [27-30]. However, the achievement is at the cost of large space occupation and complicated manipulations, such as mounting the optical cavity in a loose and soft way and adapting elaborated vibration isolation, which are almost impossible to be satisfied in field environments. For example, environmental vibrations can induce unwanted mechanical resonances, which may lead to structural deformation of the integral optical setup and deteriorate the vacuum of the chamber in which the cavity is mounted. Besides, such mechanical resonances also contribute to laser frequency noise [32]. Therefore, in addition to the pursuit of compactness and transportability, the development of ultra-stable lasers with high vibration resistance have attracted great attentions [31-41]. An important solution to avoid these vibration-induced mechanical resonances and resultant adverse effects is to characterize and increase the mechanical resonance frequency. B. Argence et al. demonstrated a full industrial engineering model of FP cavity assembly with its simulated first mechanical resonance reaching ~300 Hz, which was considered quite high considering the mass of the system [34]. D. R. Leibrandt et al. adapted the finite element analysis (FEA) method to evaluate the mechanic resonance frequency induced by the rigid-body motion [32]. Recently, G. Xu et al. further proposed the guidelines to achieve a high mechanic resonance frequency and demonstrated a trans-portable optical cavity with its first mechanic resonance significantly improved to 681 Hz [33]. However, the anti-vibration property of the whole ultra-stable laser system has not been fully investigated and optimized yet.

In this paper, the vibration analyzing model of the whole laser system including the ultra-stable optical cavity is established to seek out the weakest position, so as to meet the high anti-vibration performance by optimizing the structure. The developed laser system has a mass of ~30 kg and an integrated volume of 400 mm $\times$ 450 mm $\times$ 280 mm. Based on the model, the first-order mechanic resonance frequency of the integrated setup based on the optical cavity demonstrated in Ref. [33] was simulated to be ~170 Hz, which is mainly limited by the ion pump. Through structural optimization, it could be almost doubled to ~317.8 Hz. Via the vibration test [42], the first-order resonance frequency of the whole system was measured as ~342 Hz, whose consistence with the simulation well proved the validity of our model. The strong robustness to environmental vibrations of the developed sub-Hertz ultra-stable laser was tested by evaluating its error signal performance after a 1700-km highway transport vibration experiment, which involves a 100 km of actual transport process followed by a three-dimensional continuous shaking movement for 60 min equivalent to a 1600 km of transport vibration. According to the measurements, the ultra-stable laser system exhibited a line-width of 0.57 Hz and a fractional frequency instability of ~$2.6 \times 10^{-15}$ at 1 to 10 s averaging time, which is close to the noise floor of ~$2.0 \times 10^{-15}$. The results remained unchanged before and after the vibration test without any adjustments during the process. To our knowledge, this is the longest-distance transport vibration test ever conducted for a sub-Hertz ultra-stable laser. This test confirmed that the transportable ultra-stable laser is robust enough to meet the performance requirements of various kinds of field applications including HAGTSS.

## 2. Ultra-stable laser system

As the core of the transportable ultra-stable laser, the optical cavity must be designed as insensitive to vibration as possible and properly fixed on a support system to isolate the deterioration caused by the environmental vibrations. As shown in Fig. 1 (a), a cubic optical cavity with a length of 50 mm (L) is utilized. The three-dimensional schematic diagram of the optical cavity is provided in Fig. 1 (a), with a cutoff depth of approximately 4.2 mm at the eight vertices of the cubic optical cavity (toward the center). Both the spacer and the mirror substrates were constructed using standard-grade ULE glass. The curvature radiuses of the two mirrors were infinite (flat) and 500 mm, respectively. Both mirrors had a diameter of 25.4 mm, a thickness of 6.3 mm, and were high-reflectivity-coated for light at 1550 nm. The line-width of the optical cavity was measured to be 5.9 kHz, corresponding to a finesse of ~508000. Its physical picture on the support system is shown in Fig. 1 (b). Similar to that demonstrated in our previous work [33], the support structure of the optical cavity includes two thermal shields, a bracket, and a vacuum flange, all of which are attached to each other by four stainless steel screws. The gaskets made of PEEK were installed between them to isolate the thermal heat. The cavity was fixed to the bracket with four PEEK screws, producing a squeezing force of ~100 N at each vertex. Fig. 1(c) demonstrates the measured vibration sensitivity in the three directions of the optical cavity, with values of ~7.0/6.4/5.3 ×10$^{-11}$/g in the three directions, respectively. The cavity and its support system were placed in a vacuum chamber with a pressure of less than $1\times10^{-5}$ Pa to reduce the impact of external factors such as temperature and sound.

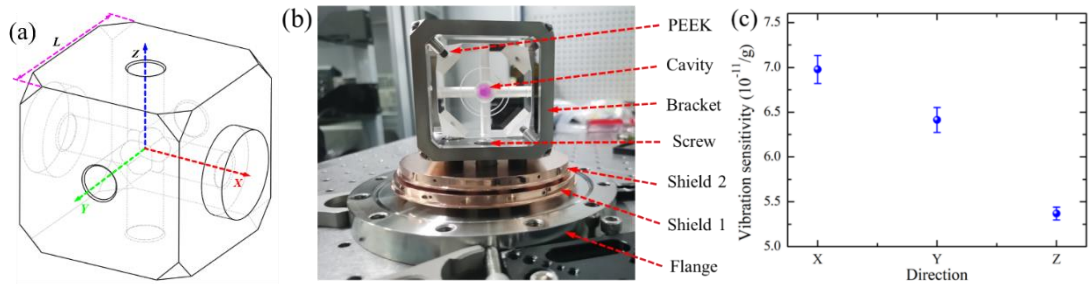

Fig. 1. (a) Three-dimensional diagram of the optical cavity. (b) Physical picture of the cavity mounted on the support system. (c) The results of measured three-dimensional vibration sensitivities.

The physical configuration of the whole ultra-stable laser system is illustrated in Fig. 2. As shown in Fig. 2(a), the internal structure of the developed transportable ultra-stable laser is made up of four parts: the fiber optical module, the spatial optical module, vacuum chamber, and the electronic module. The spatial optical module and fiber optical module were screwed on the outer wall of the vacuum chamber and reinforced with thread-fastening glue. The vacuum chamber and electronics module were laid on an 8 mm-thick aluminum base plate. As shown in Fig. 2(b), all four parts are assembled in a case of 400 mm × 450 mm × 280 mm (length × width × height) and the total mass is approximately 30 kg. The schematic diagram of the optical modules of the ultra-stable laser system is illustrated in Fig. 2(c). In the fiber optical module, a commercial fiber laser (NKT Photonics Koheras Basik E15) operating at 1550.12 nm was used as the laser source. The output beam of 20 mW was passed through a fiber-optic isolator (FOI) before being frequency-shifted by +50 MHz on an acousto-optic modulator (AOM1), which was served as the executor of fast laser frequency control. The first-order diffraction output from AOM1 was then split into two parts by a 90/10 single mode optical fiber coupler. The 90% portion is remained for further evaluation and potential applications, while the other 10% portion is utilized to realize the frequency stabilization to the optical cavity. This small portion was successively phase modulated by an electric-optical modulator (EOM) with a modulation frequency of 20 MHz. For minimizing the residual amplitude modulation (RAM) of the EOM, a fiber polarization beam splitter (FPBS) with the extinction ratio of 50 dB was put in front of it. In the spatial optical module, the output beam from the EOM (about 100 μW) laser beam was finely guided into the cavity. A homemade spatial modulator was designed to ensure the coupling efficiency of more than 20% for the laser beam into the cavity. The appropriate optical power was adjusted using a combination of a half-wave plate (HWP) and a polarization beam splitter (PBS) before the laser entered the optical cavity. The laser reflected from the surface of the optical cavity mirror was steered to a photo-detector (PD) using a combination of a quarter-wave plate (QWP) and a PBS, which was also used to prevent the laser from returning along the original path. The beat signal between the laser carrier and the modulation sidebands was detected using a PD and sent to the electronics module. This beat frequency and the signal used to drive the EOM were mixed using a double-balanced mixer (DBM) to obtain the error signal, which was then sent to Servo2 (to control the voltage of the laser piezoelectric transducer (PZT)) and Servo1 (to control the AOM driving frequency). The total delay time for the AOM driver and PD was less than 2μs, which corresponds to a servo bandwidth of 500 kHz.

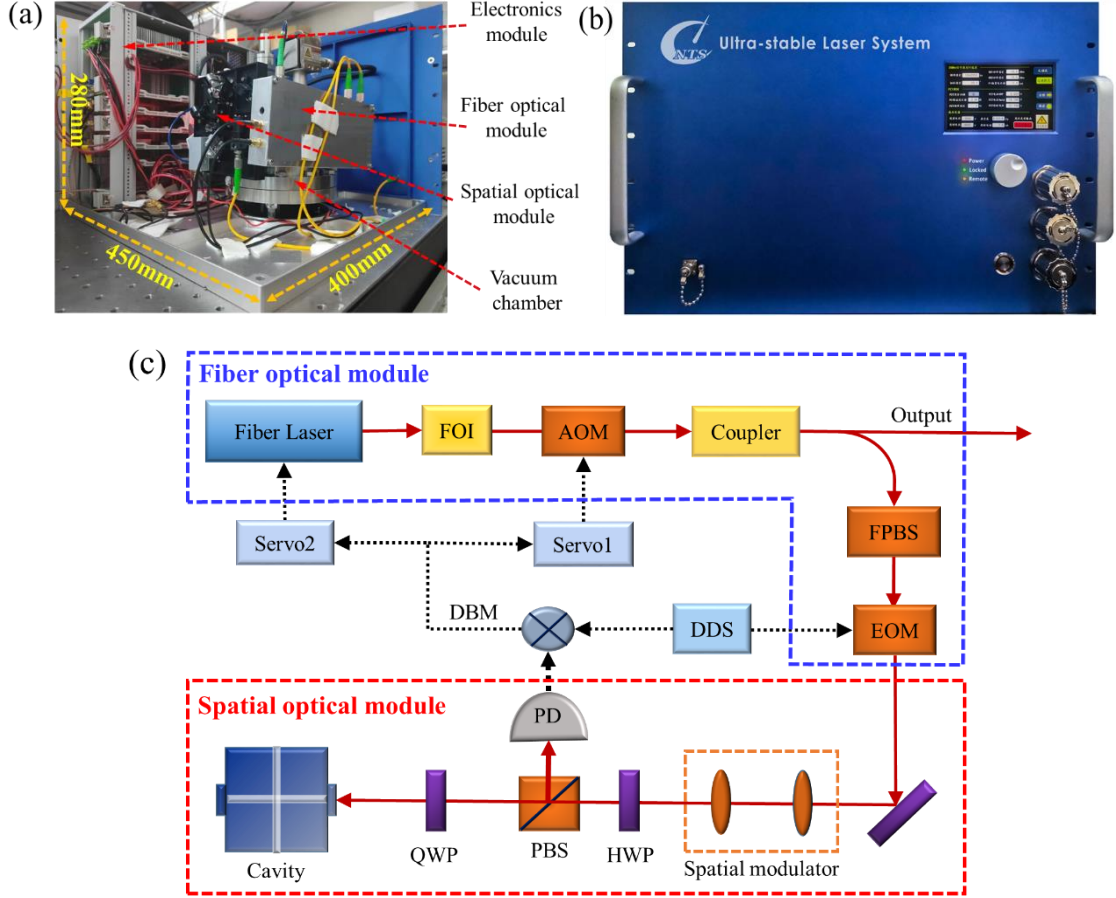

Fig. 2. (a) Internal structure of the ultra-stable laser. (b) A photograph of the physical system. (c) Its schematic diagram. FOI: Fiber-optic isolator, AOM: Fiber acoustic-optic modulator, FPBS: Optical fiber polarization beam splitter, EOM: Fiber electro-optic modulator, DDS: Direct digital synthe-sizer. PD: Photo-detector, DBM: Double-balanced mixer, HWP: Half-wave plate, PBS: Polarization beam splitter, QWP: Quarter-wave plate.

## 3. Vibration testing

To avoid potential damage caused by the complex mechanical vibrations in the working environment, the ultra-stable laser requires a higher first-order resonance frequency and improved robustness to the vibrations [32, 33]. Thus, a series of experiments and simulations were conducted to develop guidelines for designing a structure to meet these criteria [32-34]. We initially intended to design the physical part of the whole transportable ultra-stable laser similar to that of our previous work [33]. However, the FEA results show that the resonance frequency of this part is mainly limited by the ion pump and the vacuum valve, as shown in Table 1, where the red part indicates the position of the maximum resultant amplitude (AMPRES) caused by the resonance. Table 1 clearly shows that the 1st, 2nd and 4th order resonance frequencies are determined by the ion pump, and the 3rd and 5th order resonance frequencies are determined by the vacuum valve. The simulation results of the resonance frequency and the AMPRES are shown in Fig. 3 (b), respectively, with the red dotted line and the blue dotted line. The value of the first-order resonance frequency is ~147 Hz and its corresponding AMPRES is ~1.9 mm. Meanwhile, the simulation results in the first five modes show that the maximum AMPRES is ~2.6 mm. In a practical operating environment, a low first-order resonance frequency is negative for the ultra-stable laser performance, which is likely to induce the degradation of laser frequency noise [22], while the amplitude variation of the ion pump and the vacuum valve may also result in vacuum leakage. Based on the above model, the positions of the ion pump and vacuum valve were redesigned using FEA to maximize the first-order resonance frequency and reduce the synthetic amplitude variation. The improved model is shown in Fig. 3 (a), both the vacuum valve and the ion pump are located at the top of the vacuum chamber near the center. The first five order resonant models are shown in Table 1, the first two order resonant frequencies are still limited by the ion pump, and the other order resonance frequencies are limited by the support bracket of the optical cavity. The values of the resonance frequencies and the AMPRES are shown by the red and blue solid lines in Fig. 3 (b), respectively. The value of the first-order resonance frequency is raised to 317.8 Hz, which is approximately twice that of the previous model, and the maximum AMPRES is dropped from ~2.6 mm to ~1.3 mm. Through the above results, it can be seen that the vibration-resistant of the improved model has a significant improvement.

Table 1. The first five vibrational mode shapes before and after redesign

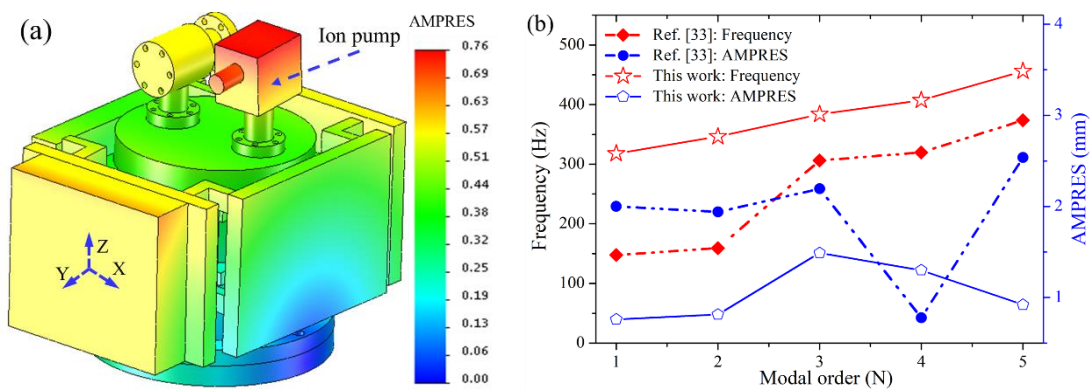

Fig. 3. Simulation of the physical part, including (a) a diagram of the FEA process in the Z direction after redesign and (b) Comparison of resonance frequency and AMPRES before and after redesign.

The ion pump is the core component to maintain the vacuum system, its damage can lead to vacuum deterioration. Therefore, the vibration characteristics on the ion pump are further analyzed to facilitate the weakening of the resonance effects by appropriate means. To better understand the distribution of vibration intensity in the ion pump, the vibration acceleration power spectral density (PSD) was simulated using FEA to mimic standard highway transportation. Simulated results in three directions are represented by the gray solid line (Z-axis), blue dotted line (X-axis), and red dashed line (Y-axis) in Fig. 4. It is evident from the figure that the intensity of the first-order resonance frequency is 0.42 $g^2$/Hz in the Z-direction, which is weakest among all directions. The first-order and second-order resonance intensities in the Y-direction are the largest, with intensities of 20.4 $g^2$/Hz and 9.7 $g^2$/Hz, respectively. These intensities in the X-direction reached 6.4 $g^2$/Hz and 1.7 $g^2$/Hz, respectively. The black line represents the acceleration PSD of the excitation source, with a test sweep frequency domain of 10–500 Hz. The acceleration PSD intensity was 0.015 $g^2$/Hz within 40 Hz and decreased gradually with a variation coefficient of 3 dB/oct from 40 Hz to 500 Hz [42].

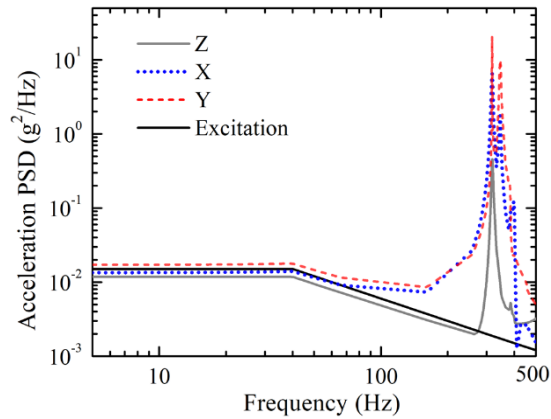

Fig. 4. Dynamic simulation of the physical part in three directions.

As shown in Fig. 5, to verify the vibration reliability of this ultra-stable laser, its robustness was tested in two phases, including 1) transportation by car from the National Time Service Center (NTSC) of the Chinese Academy of Sciences to the Aircraft Strength Research Institute in China (ASRI), a roundtrip distance of 100 km, and 2) continuous vibration on a shaker table for 60 min to simulate ~1600 km of highway. In the second phase, the system was fixed to a vibration table using two clamps and four long screw rods, as shown by an inset in Fig. 5 (a). Three accelerometers were used to measure the excitation and the response. The first accelerometer was located on the top of the vibration table and fixed with screws to measure the excitation. The others were located on the top of the vacuum chamber and ion pump, respectively, to monitor the vibration response, as illustrated in Fig. 5 (b). Signal wires for the second and third accelerometers passed through a small hole in the shell of the ultra-stable laser housing. All signal wires were fixed with anti-static stickers to reduce the influence of vibrations.

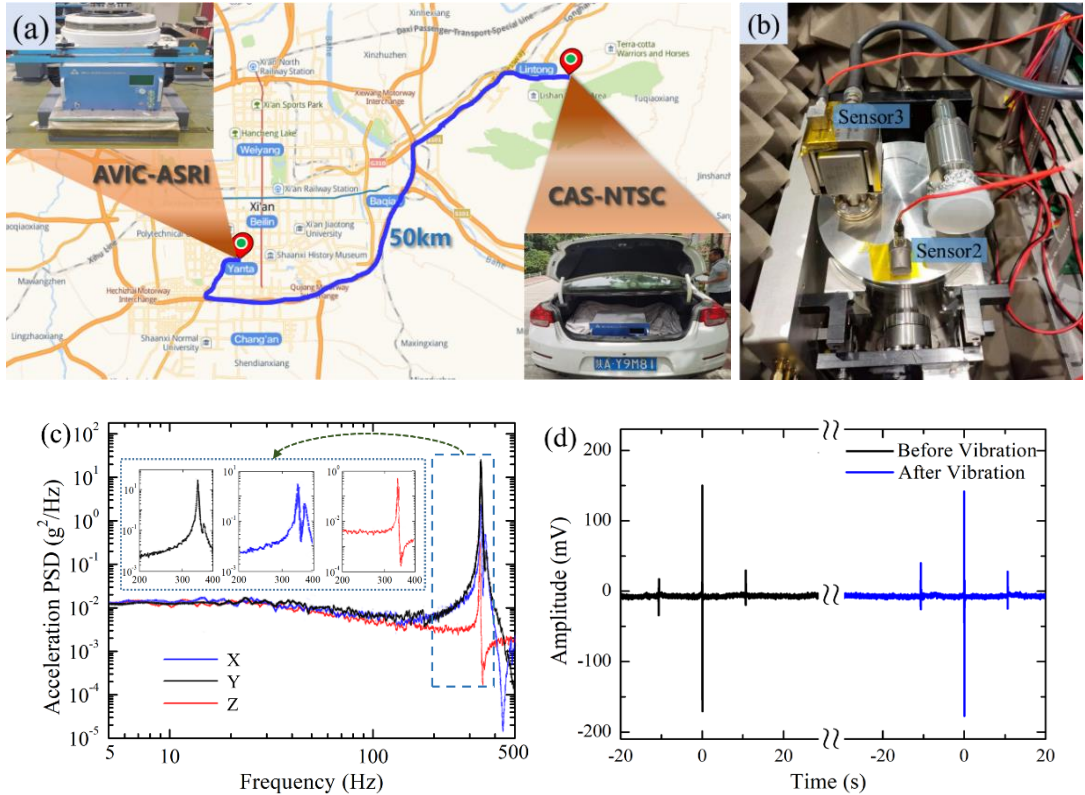

Fig. 5. Vibration tests and corresponding results, including (a) the transportation route and measure-ment site, (b) position distributions for acceleration, (c) test results from the third accelerometer in three directions, and (d) the error signals before and after vibration.

Table 2. A comparison of experimental and simulated modal results (Units – Frequency: Hz, Intensity: g2/Hz).

|  | Z-axis | | X-axis | | | | Y-axis | | | |
|---|---|---|---|---|---|---|---|---|---|---|
|  | First-order | | First-order | | Second-order | | First-order | | Second-order | |
|  | Frequency | Intensity | Frequency | Intensity | Frequency | Intensity | Frequency | Intensity | Frequency | Intensity |
| **FEA** | 317.8 | 0.42 | 317.9 | 6.4 | 346.1 | 1.7 | 317.8 | 20.4 | 346.1 | 9.7 |
| **Experiment** | 342.0 | 0.49 | 342.0 | 3.0 | 356.0 | 0.54 | 341.0 | 32.0 | 360.0 | 0.25 |

The results of the acceleration PSD are shown for the third accelerometer in Fig. 5(c). No detailed discussion is provided for the second one since the measurement results are similar to those of the excitation source. The red line in the figure represents the acceleration PSD measured along the Z-axis, which indicates the first-order resonance frequency with a value of 342 Hz and an intensity of 0.49 $g^2$/Hz. The acceleration PSD along the X-axis is represented by the blue line, with the intensities of 3.0 $g^2$/Hz and 0.54 $g^2$/Hz at corresponding resonance frequencies of 342 Hz and 356 Hz. The black line displays the acceleration PSD along the Y-axis, revealing an intensity of 32.0 $g^2$/Hz (the largest among all measured values) at the resonance frequency of 341 Hz. It is evident from Table 2 that the frequencies and intensities of resonance points measured in these three directions were consistent with simulated results. There is also a discrepancy between the FEA simulations and measured results for the first and second-order resonance peaks in all three directions. Frequency differences were approximately 8% and 3%, while differences in intensity were less than 3 dB and 16 dB, respectively. These results suggest we should be cautious in preventing resonance from damaging the ion pump and deteriorating the laser frequency noise PSD in the vibration range near 500 Hz, particularly for horizontal vibrations that can cause vacuum leakage.

Variations in the optical path can change the amplitude of the error signal, resulting in an inability to lock the laser. Since the error signal is actually the heart of the PDH technique [43], testing changes in the error signal before and after vibration is the most convenient and effective way to assert the robustness of the ultra-stable laser. The measured signals are represented in Fig. 5(d). It is evident the error signal after the vibration test (blue line) was consistent with the signal before the vibration test (black line), indicating the performance of the transportable ultra-stable laser to be extremely reliable. Note that, after the vibration test, the error signal (blue line) was measured without any adjustments.

4. Performance evaluation

We directly measured the line-width for the transportable ultra-stable laser by comparing it with another one in the laboratory. The beat note was down-converted from 726 MHz to 50 kHz and recorded using a fast Fourier transform (FFT) analyzer. The single measurement time was 2 s to allow for a sufficiently high frequency resolution bandwidth (RBW) of 0.5 Hz. The results for 100 groups, measured over the course of one hour, are shown in Fig. 6 (a). Each group of spectra was fitted with a Lorentz function to obtain the line-width of the beat note. The top half shows the probability for a line-width of less than 1 Hz to be 85%, measured by integration. The bottom half indicates the median of the line-width distribution to be ~0.78 Hz, with a probability of ~82% from 0.6 Hz to 1.0 Hz. The most probable line-width for the beat note was ~0.72 Hz. Fig. 6 (b) shows a typical line-width of 0.57 Hz (red line), the narrowest beat note spectrum we observed.

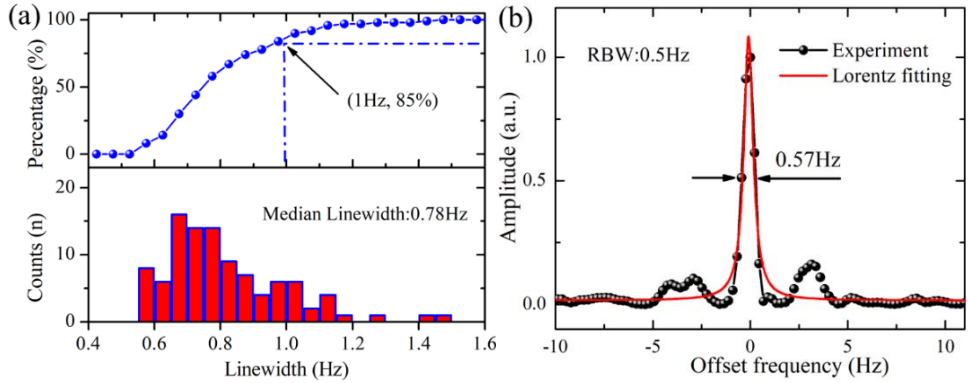

Fig. 6. Laser line-width measurements. (a) The line-width distribution of 100 groups. (b) The spectrum of beat note.

A frequency drift of ~50 kHz was observed between the two ultra-stable lasers over the course of 12 hours, as measured by a frequency counter working in Λ-mode (Agilent 53230a). As seen in Fig. 7 (a), a linear frequency drift of 1.5 Hz/s was observed during the first 9 hours and a value of 1.8 Hz/s was seen from the 9th hour to the 12th hour. The fractional frequency instability is represented by the blue line shown in Fig. 7 (b). After removing a linear drift of ~1.5 Hz/s, the typical frequency instability from the 2nd hour to the 3rd hour was ~$2.6 \times 10^{-15}$ at an averaging time of 1–10 s, as indicated by the blue line in the figure. This value is very close to the total noise floor of ~$2.0 \times 10^{-15}$, including a thermal noise limit of ~$1.8 \times 10^{-15}$ (dashed red line) and a residual amplitude noise of ~$8 \times 10^{-16}$ (solid pink line) at an averaging time of 1–10 s. For short averaging times, this increase in in-stability is due to vibrations and acoustic noise. At long averaging times, temperature fluctuations are the primary contributor to fractional frequency instability.

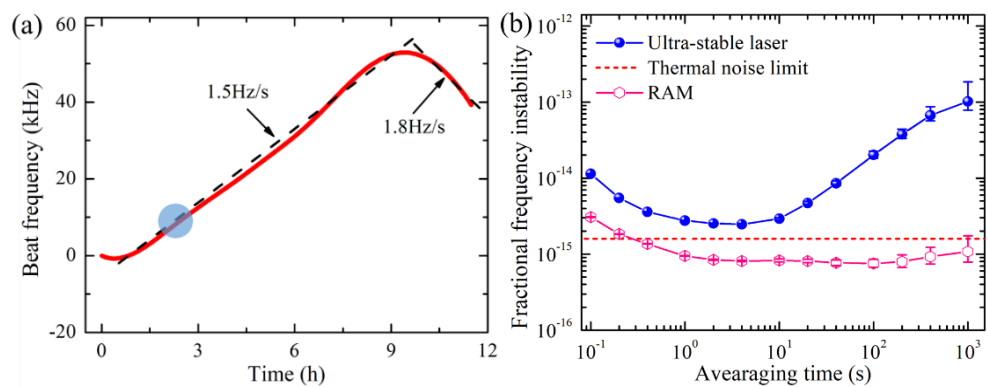

Fig. 7. Measurements of the fractional frequency instability. (a) The laser frequency drift. (b) Frequency instability.

## 5. Conclusion

We have demonstrated a sub-hertz transportable ultra-stable laser using the longest distance practical transportation test to date. A dynamic analytical model of this ultra-stable laser was established and studied in detail using both FEA simulations and physical experiments, to reduce the effects of vibrations. The resonance frequencies for the optimized system, determined by FEA, agreed well with experimentally measured results. The performance of the ultra-stable laser remained nearly unchanged after more than 1700 km transportation, verifying its robustness. In addition, the line-width of this transportable ultra-stable laser was 0.57 Hz, with a fractional frequency instability of $\sim 2.6 \times 10^{-15}$ at 1–10 s, approaching the noise floor of $\sim 2.0 \times 10^{-15}$. These results suggest the proposed system to be a good candidate for optical frequency transfer in the HAGTSS. The design concept proposed in this paper could also be helpful for ultra-stable lasers operating in more complex environments.

**CRediT authorship contribution statement**

**Dongdong Jiao**: Formal analysis, Experiment, Investigation, Software, Writing-original draft, Writing - review & editing. **Xue Deng**: Conceptualization, Resources, Writing – review & editing. **Guanjun Xu**: Conceptualization, Software, Writing – review & editing. **Jing Gao**: Resources. **Linbo Zhang**: Resources. **Tao Liu**: Supervision. **Ruifang Dong**: Supervision. **Shougang Zhang**: Supervision.

**Declaration of Competing Interest**

The authors declare that they have no known competing financial interests or personal relationships that could have appeared to influence the work reported in this paper.

**Acknowledgments**

This project wss supported by the National Natural Science Foundation of China (NSFC) (12103059), Xi'an Science and Technology Bureau (E019XK1S04), and the Youth Innovation Promotion Association of the Chinese Academy of Science (1188000XGJ).

**References**

[1] C. W. Chou, D. B. Hume, J. C. J. Koelemeij, D. J. Wineland, T. Rosenband. Frequency comparison of two high accuracy Al$^+$ optical clocks, Phys. Rev. Lett. 104 (2010) 070802, https://doi.org/10.1103/PhysRevLett.104.070802.

[2] T. L. Nicholson, S. L. Campbell, R. B. Hutson, G.E. Marti, B. J. Bloom, R. L. Mcnally, W. Zhang, M. D. Barrett, M. S. Safronova, F. Strouse, Systematic evaluation of an atomic clock at $2 \times 10^{-18}$ total uncertainty, Nat. Commun. 6 (2015) 6896, https://doi.org/10.1038/ncomms7896.

[3] N. Huntemann, C. Sanner, B. Lipphardt, C. Tamm, E. Peik, Single-ion atomic clock with $3 \times 10^{-18}$ systematic uncertainty, Phys. Rev. Lett. 116 (2016) 063001, https://doi.org/10.1103/PhysRevLett.116.063001.

[4] M. Schioppo, R. C. Brown, W. F. Mcgrew, N. Hinkley, R. J. Fasano, K Beloy, T. H. Yoon, G. Milani, D. Nicolodi, J. A. Sherman, Ultra-stable optical clock with two cold-atom ensembles, Nat. Photon. 11 (2017) 48–52, https://doi.org/10.1038/NPHOTON.2016.231.

[5] T. Takano, M. Takamoto, I. Ushijima, N. Ohmae, T. Akatsuka, A. Yamaguchi, Y. Kuroishi, H. Munekane, B. Miyahara, H. Katori, Demonstration of $4.8 \times 10^{-17}$ stability at 1 s for two independent optical clocks. Nat. Photon. 13 (10) (2019) 714–719, https://doi.org/10.1038/s41566-019-0493-4.

[6] R. M. Godun, R. M. Godun, P. B. R. Nisbet-Jones, J. M. Jones, S. A. King, L. A. M. Johnson, H. S. Margolis, K. Szymaniec, S. N. Lea, K. Bongs, P. Gill, Frequency ratio of two optical clock transitions in $^{171}$Yb$^+$ and constraints on the time variation of fundamental constants, Phys. Rev. Lett. 113 (2014) 210801, https://doi.org/10.1103/PHYSREVLETT.113.210801.

[7] N. Huntemann, B. Lipphardt, C. Tamm, V. Gerginov, S. Weyers, E. Peik, Improved limit on a temporal variation of mp/me from comparisons of Yb$^+$ and Cs atomic clocks, Phys. Rev. Lett. 113 (2014) 210802, https://doi.org/10.1103/PhysRevLett.113.210802.

[8] P. Delva, J. Lodewyck, S. Bilicki, E. Bookjans, G. Vallet, R. Le Targat, P. E. Pottie, C. Guerlin, F. Meynadier, C. Le Poncin-Lafitte, O. Lopez, A. Amy-Klein, W. K. Lee, N. Quintin, C. Lisdat, A. Al-Masoudi, S. Doerscher, C. Grebing, G. Grosche, A. Kuhl, S. Raupach, U. Sterr, I. R. Hill, R. Hobson, W. Bowden, J. Kronjager, G. Marra, A. Rolland, F. N. Baynes, H. S. Margolis, P. Gill, Test of special relativity using a fiber network of optical clocks, Phys. Rev. Lett. 118 (2017) 221102, https://doi.org/10.1103/PhysRevLett.118.221102.

[9] P. Wcisło, P. Ablewski, K. Beloy, S. Bilicki, M. Bober, R. Brown, R. Fasano, R. Ciuryło, H. Hachisu, T. Ido, J. Lodewyck, A. Ludlow, W. McGrew, P. Morzyński, D. Nicolodi, M. Schioppo, M. Sekido, R. Le Targa, P. Wolf, X. Zhang, B. Zjawin, M. Zawada, New bounds on dark matter coupling from a global network of optical atomic clocks, Sci. Adv. 4 (2018) 1–7, https://doi.org/10.1126/sciadv.aau4869.

[10] B.M. Roberts, P. Delva, A. Al-Masoudi, A. Amy-Klein, C. Baerentsen, C.F.A. Baynham, E. Benkler, S. Bilicki, S. Bize, W. Bowden, J. Calvert, V. Cambier, E. Cantin, E.A. Curtis, S. Dörscher, M. Favier, F. Frank, P. Gill, R.M. Godun, G. Grosche, C. Guo, A. Hees, I.R. Hill, R. Hobson, N. Huntemann, J. Kronjäger, S. Koke, A. Kuhl, R. Lange, T. Legero, B. Lipphardt, C. Lisdat, J. Lodewyck, O. Lopez, H.S. Margolis, H. Álvarez-Martínez, F. Meynadier, F. Ozimek, E. Peik, P.-E Pottie, N. Quintin, C. Sanner, L. de Sarlo, M. Schioppo, R. Schwarz, A. Silva, U. Sterr, C. Tamm, R. L. Targat, P. Tuckey, G. Vallet, T. Waterholter, D. Xu, P. Wolf, Search for transient variations of the fine structure constant and dark matter using fiber-linked optical atomic clocks, N. J. Phys. 22 (2020) 093010, https://doi.org/10.1088/1367-2630/abaace

[11] C. Lisdat, G. Grosche, N. Quintin, C. Shi, S.M.F. Raupach, C. Grebing, D. Nicolodi, F. Stefani, A. Al-Masoudi, S. Dörscher, S. Häfner, J.-L. Robyr, N. Chiodo, S. Bilicki, E. Bookjans, A. Koczwara, S. Koke, A. Kuhl, F. Wiotte, F. Meynadier, E. Camisard, M. Abgrall, M. Lours, T. Legero, H. Schnatz, U. Sterr, H. Denker, C. Chardonnet, Y. Le Coq, G. Santarelli, A. Amy-Klein, R. Le Targat, J. Lodewyck, O. Lopez, P. E. Pottie, A clock network for geodesy and fundamental science. Nat. commun. 7 (2016) 12443, https://doi.org/10.1038/ncomms12443.

[12] T. Takano, M. Takamoto, I. Ushijima, N. Ohmae, T. Akatsuka, A. Yamaguchi, Y. Kuroishi, H. Munekane, B. Miyahara, H. Katori, Geopotential measurements with synchronously linked optical lattice clocks, Nat. Photon. 10 (2016) 662–666, https://doi.org/10.1038/nphoton.2016.159.

[13] J. Grotti, S. Koller, S. Vogt, S. Häfner, U. Sterr, C. Lisdat, H. Denker, C. Voigt, L. Timmen, and A. Rolland, Geodesy and metrology with a transportable optical clock, Nat, Phys, 14 (2018) 437–441, https://doi.org/10.1038/s41567-017-0042-3.

[14] M. Takamoto, I. Ushijima, N. Ohmae, T. Yahagi, H. Katori, Test of general relativity by a pair of transportable optical lattice clocks, Nat. Photon. 14 (2020) 411–415, https://doi.org/10.1038/s41566-020-0619-8.

[15] F. Riehle, Towards a redefinition of the second based on optical atomic clocks, C. R. Phys. 16 (5) (2015) 506–515, https://doi.org/10.1016/j.crhy.2015.03.012.

[16] S. Bize, The unit of time: Present and future directions, C. R. Phys. 20 (1-2) (2019) 153–168, https://doi.org/10.1016/j.crhy.2019.02.002.

[17] W. F. McGrew, X. Zhang, R. J. Fasano, S. A. Schäffer, K. Beloy, D. Nicolodi, R. C. Brown, N. Hinkley, G. Milani, M. Schioppo, T. H. Yoon, A. D. Ludlow, Atomic clock performance enabling geodesy below the centimetre level, Nat. 564 (2018) 87–90, https://doi.org/ 10.1038/s41586-018-0738-2.

[18] A. D. Ludlow, An optical clock to go, Nat. Phys. 14 (2018) 431–432, https://doi.org/10.1038/s41567-018-0047-6.

[19] M. Schioppo, J. Kronjäger, A. Silva, R. Ilieva, J. W. Paterson, C. F. A. Baynham, W. Bowden, I. R. Hill, R. Hobson, A. Vianello, M. Dovale-Álvarez, R. A. Williams, G. Marra, H. S. Margolis, A. Amy-Klein, O. Lopez, E. Cantin, H. Álvarez-Martínez, R. Le Targat, P. E. Pottie, N. Quintin, T. Legero, S. Häfner, U. Sterr, R. Schwarz, S. Dörscher, C. Lisdat, S. Koke, A. Kuhl, T. Waterholter, E. Benkler, G. Grosche, Comparing ultrastable lasers at $7 \times 10^{-17}$ fractional frequency instability through a 2220 km optical fibre network, Nat. coummun. 13 (2022) 212, https://doi.org/10.1038/s41467-021-27884-3.

[20] S. Droste, F. Ozimek, Th. Udem, K. Predehl, T. W. Hänsch, H. Schnatz, G. Grosche, R. Holzwarth, Optical-frequency Transfer over a Single-Span 1840 km Fiber Link, Phys. Rev. Lett. 111 (2013) 110801, https://doi.org/10.1103/PhysRevLett.111.110801.

[21] D. Calonico, E. K. Bertacco, C. E. Calosso, C. Clivati, G. A. Costanzo, M. Frittelli, A. Godone, A. Mura, N. Poli, D. V. Sutyrin, G. Tino, M. E. Zucco, F. Levi, High-accuracy coherent optical frequency transfer over a doubled 642-km fiber link, Appl. Phys. B 117 (2014) 979-986, https://doi.org/10.1007/s00340-014-5917-8.

[22] J. Vojtěch, J. Radil, V. Smotlacha, R. Velc, P. Krehlik, Ł. Śliwczyński, M. Campanella, D.Calonico, C. Clivati, F. Levi, O. Číp, S. Rerucha, R. Holzwarth, M. Lessing, S. Saint-Jalm, F. Camargo, B. Desruelle, J. Lautier-Gaud, E. L. English, J. Kronjäger, P. Whibberley, E. Bookjans, P. E. Pottie, P. Tuckey, T. Müller, J. Štefl, M. Šteflová, P. Nogas, R. Urbaniak, A. Binczewski, W. Bogacki, K. Turza, G. Grosche, H. Schnatz, E. Camisard, N. Quintin, J. Diaz, E. Ros, T. García, A. Galardini, A. Seeds, Z. Yang, A. Amy-Klein, The 2020 Project CLONETS: Clock Services over Optical-fibre Networks in Europe. CLEO: Applications and Technology, Lasers and Electro-Optics, May 2018, https://doi.org/10.1364/CLEO_AT.2018.JW2A.137.

[23] Clock Network Services Strategy and innovation for clock services over optical-fibre networks. https://www.clonets.eu/.

[24] S. Zhang, "Precise generation and transfer of time and frequency in NTSC," ESA Topical Team on Geodesy and Clocks in Space, Oct. 2019, https://www.asg.ed.tum.de/fileadmin/w00cip/fesg/aces/21.pdf.

[25] 70th Anniversary of Beijing Time-High Accurate Ground-based Time Service System, http://english.ntsc.cas.cn/newsroom/sr/202108/t20210811_277988.html.

[26] B. Parker, S. Webster, S. Lea, P. Gill, P. Bayvel, Development of an ultra-stable laser in the 1.5 μm band for optical frequency transfer over optical fibre, Eftf- European Frequency & Time Forum. IEEE, Apr. 2010, https://ieeexplore.ieee.org/document/6533718

[27] S. Häfner, S. Falke, C. Grebing, S. Vogt, T. Legero, M. Merimaa, C. Lisdat, U. Sterr, $8 \times 10^{-17}$ fractional laser frequency instability with a long room-temperature cavity, Opt. Lett. 40 (2015) 2112–2115, https://doi.org/10.1364/OL.40.002112.

[28] W. Zhang, J. M. Robinson, L. Sonderhouse, E. Oelker, C. Benko, J. L. Hall, T. Legero, D. G. Matei, F. Riehle, U. Sterr, J. Ye, An ultrastable Silicon Cavity in a Continuously Operating Closed-Cycle Cryostat at 4 K, Phys .Rev. Lett. 119 (2017) 243601, https://doi.org/10.1103/PhysRevLett.119.243601.

[29] D. G. Matei, T. Legero, S. Häfner, C. Grebing, R. Weyrich, W. Zhang, L. Sonderhouse, J. M. Robinson, J. Ye, F. Riehle, U. Sterr, 1.5 μm Lasers with Sub-10 mHz Linewidth, Phys. Rev. Lett. 118 (2017) 263202, https://doi.org/10.1103/PhysRevLett.118.263202.

[30] D. Jiao, J. Gao, X. Deng. G. Xu, J. Liu, T. Liu, R. Dong, S. Zhang, Sub-Hertz frequency stabilization of 1.55 μm laser on Higher order HGmn, Opt. Commun. 463 (2020) 125460, https://doi.org/10.1016/j.optcom.2020.125460.

[31] S. Herbers, Transportable ultra-stable laser system with an instability down to $10^{-16}$, (Ph.D. thesis). Gottfried Wilhelm Leibniz Universität Hannover 2021.p:46.

[32] D. R. Leibrandt, J. C. Bergquist, T. Rosenband, Cavity-stabilized laser with acceleration sensitivity below $10^{-12}$ $g^{-1}$, Phys. Rev. A. 87 (2013) 023829, https://doi.org/10.1103/PhysRevA.87.023829.

[33] G. Xu, D. Jiao. L. Chen, L. Zhang, R. Dong, T. Liu J. Wang, Vibration Modes of a Transportable Optical Cavity, Opt. Express 29 (15) (2021) 24264, https://doi.org/10.1364/OE.422182.

[34] B. Argence, B. Argence, E. Prevost, T. Lévèque, R. Le Goff, S. Bize, P. Lemonde, G. Santarelli, Prototype of an ultra-stable optical cavity for space applications, Opt. Express 20 (23) (2012) 25409-25420, https://doi.org/10.1364/OE.20.025409.

[35] D. Świerad, S. Häfner, S.Vogt, B. Venon, D. Holleville, S. Bize, A. Kulosa, S. Bode, Y. Singh, K. Bongs, E. M. Rasel, J. Lodewyck, R. L. Targat, C. Lisdat, U. Sterr, Ultra-stable clock laser system development towards space applications, Sci. Rep. 6 (2016) 33973, https://doi.org/10.1038/srep33973.

[36] Q. Chen, A. Nevsky, M. Cardace, S. Schiller, T. Legero, S. Häfner, A. Uhde, U. Sterr, A compact, robust, and transportable ultra-stable laser with a fractional frequency instability of $1 \times 10^{-15}$, Rev. Sci. Instrum. 85 (2014) 113107, https://doi.org/10.1063/1.4898334

[37] S. Webster, and P. Gill, Force-insensitive optical cavity, Opt. Lett. 36 (18) (2011) 3572, https://doi.org/10.1364/OL.36.003572

[38] X. Chen, Y. Jiang, B. Li, H. Yu, H. Jiang, T. Wang, Laser frequency instability of $6 \times 10^{-16}$ using 10-cm-long cavities on a cubic spacer, Chin. Opt. Lett. 18 (3) (2020) 030201, https://doi.org/10.1364/COL.18.030201.

[39] S. Häfner, S. Herbers, S. Vogt, C. Lisdat, U. Sterr, Transportable interrogation laser system with an instability of mod σy=3 $\times 10^{-16}$, Opt. Express 28 (11) (2020) 16407-16416, https://doi.org/10.1364/OE.390105.

[40] S. Herbers, S. Häfner, S. Dörscher, T. Lücke, U. Sterr, C. Lisdat, A transportable clock laser system with an instability of mod σy=1.6 $\times 10^{-16}$, Opt. Lett. 47 (2022) 5441-5444, https://doi.org/10.1364/OL.470984.

[41] S. Wang, J. Cao, J. Yuan, D. Liu, H. Shu, X. Huang, Integrated multiple wavelength stabilization on a multi-channel cavity for a transportable optical clock, Opt. Express. 28 (11) (2020)16407-16416, https://doi.org/10.1364/OE.383115

[42] GJB150. 1A-2009, Laboratory environmental test methods of military materiel, China, 2009.

[43] E. D. Black, An introduction to Pound Drever Hall laser frequency stabilization, Amer. J. Phys. 69 (1) (2000) 79–87, https://doi.org/10.1119/1.1286663.